\documentclass[aip,rsi,amsmath,amssymb,reprint]{revtex4-1}

\usepackage{graphicx}
\usepackage{siunitx}
\usepackage{dcolumn}
\usepackage{amsmath}
\usepackage{amssymb}
\usepackage[english]{babel}

\begin{document}

\title{Trap-integrated fluorescence detection based on silicon photomultipliers in a cryogenic Penning trap}

\author{M. Wiesinger}
\affiliation{Max-Planck-Institut für Kernphysik, Saupfercheckweg 1, 69117 Heidelberg, Germany}
 
\author{F. Stuhlmann}
\affiliation{Institut für Physik, Johannes Gutenberg-Universität Mainz, Staudingerweg 7, 55128 Mainz, Germany}

\author{M. Bohman}
\affiliation{Max-Planck-Institut für Kernphysik, Saupfercheckweg 1, 69117 Heidelberg, Germany}

\author{P. Micke}
\affiliation{Max-Planck-Institut für Kernphysik, Saupfercheckweg 1, 69117 Heidelberg, Germany}
\affiliation{CERN, Esplanade des Particules 1, 1217 Meyrin, Switzerland}

\author{C. Will}
\affiliation{Max-Planck-Institut für Kernphysik, Saupfercheckweg 1, 69117 Heidelberg, Germany}

\author{H. Yildiz}
\affiliation{Institut für Physik, Johannes Gutenberg-Universität Mainz, Staudingerweg 7, 55128 Mainz, Germany}

\author{F. Abbass}
\affiliation{Institut für Physik, Johannes Gutenberg-Universität Mainz, Staudingerweg 7, 55128 Mainz, Germany}

\author{B. P. Arndt}
\affiliation{Max-Planck-Institut für Kernphysik, Saupfercheckweg 1, 69117 Heidelberg, Germany}
\affiliation{GSI Helmholtzzentrum für Schwerionenforschung GmbH, Planckstraße 1, 64291 Darmstadt, Germany}
\affiliation{RIKEN, Fundamental Symmetries Laboratory, 2-1 Hirosawa, Wako, Saitama 351-0198, Japan}

\author{J. A. Devlin}
\affiliation{CERN, Esplanade des Particules 1, 1217 Meyrin, Switzerland}
\affiliation{RIKEN, Fundamental Symmetries Laboratory, 2-1 Hirosawa, Wako, Saitama 351-0198, Japan}

\author{S. Erlewein}
\affiliation{Max-Planck-Institut für Kernphysik, Saupfercheckweg 1, 69117 Heidelberg, Germany}
\affiliation{RIKEN, Fundamental Symmetries Laboratory, 2-1 Hirosawa, Wako, Saitama 351-0198, Japan}

\author{M. Fleck}
\affiliation{RIKEN, Fundamental Symmetries Laboratory, 2-1 Hirosawa, Wako, Saitama 351-0198, Japan}
\affiliation{Graduate School of Arts and Sciences, University of Tokyo, 3-8-1 Komaba, Meguro, Tokyo 153-8902, Japan}

\author{J. I. Jäger}
\affiliation{Max-Planck-Institut für Kernphysik, Saupfercheckweg 1, 69117 Heidelberg, Germany}
\affiliation{CERN, Esplanade des Particules 1, 1217 Meyrin, Switzerland}
\affiliation{RIKEN, Fundamental Symmetries Laboratory, 2-1 Hirosawa, Wako, Saitama 351-0198, Japan}

\author{B. M. Latacz}
\affiliation{CERN, Esplanade des Particules 1, 1217 Meyrin, Switzerland}
\affiliation{RIKEN, Fundamental Symmetries Laboratory, 2-1 Hirosawa, Wako, Saitama 351-0198, Japan}

\author{D. Schweitzer}
\affiliation{Institut für Physik, Johannes Gutenberg-Universität Mainz, Staudingerweg 7, 55128 Mainz, Germany}

\author{G. Umbrazunas}
\affiliation{RIKEN, Fundamental Symmetries Laboratory, 2-1 Hirosawa, Wako, Saitama 351-0198, Japan}
\affiliation{Eidgenössische Technische Hochschule Zürich, John-von-Neumann-Weg 9, 8093 Zürich, Switzerland}

\author{E. Wursten}
\affiliation{CERN, Esplanade des Particules 1, 1217 Meyrin, Switzerland}
\affiliation{RIKEN, Fundamental Symmetries Laboratory, 2-1 Hirosawa, Wako, Saitama 351-0198, Japan}

\author{K. Blaum}
\affiliation{Max-Planck-Institut für Kernphysik, Saupfercheckweg 1, 69117 Heidelberg, Germany}

\author{Y. Matsuda}
\affiliation{Graduate School of Arts and Sciences, University of Tokyo, 3-8-1 Komaba, Meguro, Tokyo 153-8902, Japan}

\author{A. Mooser}
\affiliation{Max-Planck-Institut für Kernphysik, Saupfercheckweg 1, 69117 Heidelberg, Germany}

\author{W. Quint}
\affiliation{GSI Helmholtzzentrum für Schwerionenforschung GmbH, Planckstraße 1, 64291 Darmstadt, Germany}

\author{A. Soter}
\affiliation{Eidgenössische Technische Hochschule Zürich, John-von-Neumann-Weg 9, 8093 Zürich, Switzerland}

\author{J. Walz}
\affiliation{Institut für Physik, Johannes Gutenberg-Universität Mainz, Staudingerweg 7, 55128 Mainz, Germany}
\affiliation{Helmholtz-Institut Mainz, Staudingerweg 18, 55128 Mainz, Germany}

\author{C. Smorra}
\affiliation{Institut für Physik, Johannes Gutenberg-Universität Mainz, Staudingerweg 7, 55128 Mainz, Germany}
\affiliation{RIKEN, Fundamental Symmetries Laboratory, 2-1 Hirosawa, Wako, Saitama 351-0198, Japan}

\author{S. Ulmer}
\affiliation{RIKEN, Fundamental Symmetries Laboratory, 2-1 Hirosawa, Wako, Saitama 351-0198, Japan}
\affiliation{Heinrich-Heine-Universität Düsseldorf, Universitätsstrasse 1, 40225 Düsseldorf, Germany}

\collaboration{BASE Collaboration}

\date{August 3, 2023}

\begin{abstract}
We present a fluorescence-detection system for laser-cooled \textsuperscript{9}Be\textsuperscript{+} ions based on silicon photomultipliers (SiPM) operated at \SI{4}{\kelvin} and integrated into our cryogenic \SI{1.9}{\tesla} multi-Penning-trap system. Our approach enables fluorescence detection in a hermetically-sealed cryogenic Penning-trap chamber with limited optical access, where state-of-the-art detection using a telescope and photomultipliers at room temperature would be extremely difficult. We characterize the properties of the SiPM in a cryo\-cooler at \SI{4}{\kelvin}, where we measure a dark count rate below \SI{1}{\per\second} and a detection efficiency of 2.5(3)\,\%. We further discuss the design of our cryogenic fluorescence-detection trap, and analyze the performance of our detection system by fluorescence spectroscopy of \textsuperscript{9}Be\textsuperscript{+} ion clouds during several runs of our experiment.
\end{abstract}

\maketitle

\section{Introduction}

Detection of fluorescence photons is an essential tool in experiments with laser-cooled trapped ions.
In early experiments with single trapped ions it allowed the first observation of quantum jumps~\cite{Nagourney1986,Sauter1986,Bergquist1986}.
In state-of-the-art trapped-ion quantum computers it facilitates high-fidelity qubit readout~\cite{Myerson2008}.
In fundamental physics experiments it enables the application of sympathetic ground-state cooling and quantum logic spectroscopy and, therefore, the extension of laser-cooling techniques to ions without suitable laser-cooling transitions~\cite{Brewer2019,Micke2020}.
So far, all these experiments rely on collection of fluorescence light with high numerical aperture optics and detection with a photomultiplier tube or camera at room temperature.

Despite Penning traps being indispensable tools for fundamental physics experiments where high magnetic fields are essential, e.g.~for $g$-factor or mass measurements of single trapped ions \cite{Schneider2017,Smorra2017,Borchert2022,Schuessler2020,Rau2020,Schneider2022}, these experiments are usually not equipped with fluorescence detectors. Generally, optical access is at a premium because a Penning trap is usually located inside the bore of a superconducting magnet, and in most cases cooled to cryogenic temperatures. Where fluorescence detection has been used, complicated optical pathways have been required to bring the fluorescence photons to the detection system located outside the magnet bore.
Examples of such Penning-trap setups are experiments on motional ground-state cooling of calcium ions~\cite{Goodwin2016,Hrmo2019}, experiments with two-dimensional ion crystals for quantum simulation~\cite{Jordan2019}, mass measurements of heavy ions~\cite{Gutierrez2019}, and laser spectroscopy of highly-charged ions~\cite{Schmidt2017}.

In this paper, we present a fluorescence-detection system based on MicroFJ-30035-TSV silicon photomultipliers (SiPM) from \textit{onsemi}~\cite{SiPM_datasheet}, which are integrated into the electrode structure of our cryogenic Penning-trap system.
Our approach does not require an optical pathway to the outside of the magnet bore. 
This is especially useful for experiments where the Penning-trap system is enclosed in a hermetically-sealed vacuum chamber and cooled to cryogenic temperatures in order to utilize cryogenic pumping to achieve extreme-high vacuum, for instance allowing for antiproton storage times of years~\cite{Sellner2017}.
Due to their compact dimensions and expected insensitivity to magnetic fields, SiPM are ideally suited for operation in this environment.
Furthermore, it has been shown that some SiPM are also compatible with cryogenic temperatures down to \SI{4}{\kelvin}~\cite{Biroth2015,Biroth2016,Achenbach2018}.
While the dark count rate of SiPM is typically several \SI{e4}{\per\second\per\milli\meter\squared} at room temperature, at cryogenic temperatures this problem is greatly reduced leading to extremely low dark count rates below a few counts per second.
Further, it should be noted that SiPM are a relatively inexpensive commercial product, available in a variety of models, thus avoiding the development of custom-made devices.
Related approaches of trap-integrated fluorescence detection use custom micro-fabricated superconducting sensors in a cryogenic radio-frequency trap~\cite{Slichter2021} or custom chip-integrated avalanche photodiodes in a room temperature radio-frequency trap~\cite{Reens2022}.

The work on trap-integrated detection of fluorescence is inspired by our experiments on sympathetic cooling of a single proton by laser-cooled \textsuperscript{9}Be\textsuperscript{+} ions~\cite{Bohman2021}. These efforts will lead to a new cooling method for single protons and antiprotons. The final temperatures in the \si{\milli\kelvin} range will be needed for the next generation of high-precision measurements of the proton and antiproton $g$-factors~\cite{Will2022}. The newly developed trap-integrated fluorescence-detection system is compatible with the hermetically-sealed trap chamber required for these measurements.
In our experiment the fluorescence-detection system is used for the determination of the resonance frequency of the cooling transition in our \SI{1.9}{\tesla} magnetic field, for optimization of the cooling-laser parameters regarding intensity, position and polarization, and for determining the axial temperature of the trapped \textsuperscript{9}Be\textsuperscript{+} ion cloud. 
Ultimately, fluorescence-based state readout of a \textsuperscript{9}Be\textsuperscript{+} ion coupled to a proton or antiproton can be used for sympathetic cooling and implementation of quantum logic spectroscopy for Larmor and cyclotron frequency measurements on the proton or antiproton~\cite{Heinzen1990,Cornejo2021}.

In the following section we describe the design of the Penning-trap system used in our experiments, in section \ref{sec:SiPMcharacter} we characterize and compare the SiPM properties at room temperature and at \SI{4}{\kelvin}, in section \ref{sec:fluorescence_detection} we show measurements of fluorescence photon counts from a cloud of Doppler laser-cooled \textsuperscript{9}Be\textsuperscript{+} ions and determine the axial temperature of the trapped ion cloud. We summarize the results in section \ref{sec:conclusion}.

\section{Experimental setup}

The multi-Penning-trap system used in this work is designed for a future high-precision measurement of the proton $g$-factor with relative uncertainty of $10^{-11}$.
It consists of a stack of six cylindrical open-endcap Penning traps with ion transport capability between all traps. Two traps implement the double-Penning-trap technique for $g$-factor measurements~\cite{Haeffner2003,Mooser2013}. 
Two other traps are used to couple a single proton to a cloud of \textsuperscript{9}Be\textsuperscript{+} ions for sympathetic cooling~\cite{Heinzen1990}.
An initial design of these traps has been described previously~\cite{Bohman2017}.
Two new traps have been added which are used for high-resolution particle temperature measurements and particle loading through laser ablation, respectively~\cite{WiesingerPhD}.
\textsuperscript{9}Be\textsuperscript{+} ions are loaded from a beryllium foil using a single \SI{5}{\nano\second} long pulse from a frequency-doubled Nd:YAG laser at \SI{532}{\nano\meter} with \SI{0.2}{\milli\joule} to \SI{0.6}{\milli\joule} pulse energy. 
Trapped ions are detected using non-destructive image-current detection systems. To this end, one electrode in each trap is connected to a superconducting LC circuit, which also resistively cools the ions to near \SI{4}{\kelvin}~\cite{Ulmer2009,Nagahama2016}.

The trap system is enclosed in a hermetically-sealed vacuum chamber which is cooled to $\approx \SI{4}{\kelvin}$ and located inside the bore of a superconducting magnet. 
Optical and laser access is extremely limited and only possible along the axial direction through small fused-silica windows in the trap chamber. Instead of routing the fluorescence light to room temperature, which would require complicated optical pathways, we pursue trap-integrated detection of fluorescence.

\begin{figure}
    \includegraphics[width = 0.55\linewidth]{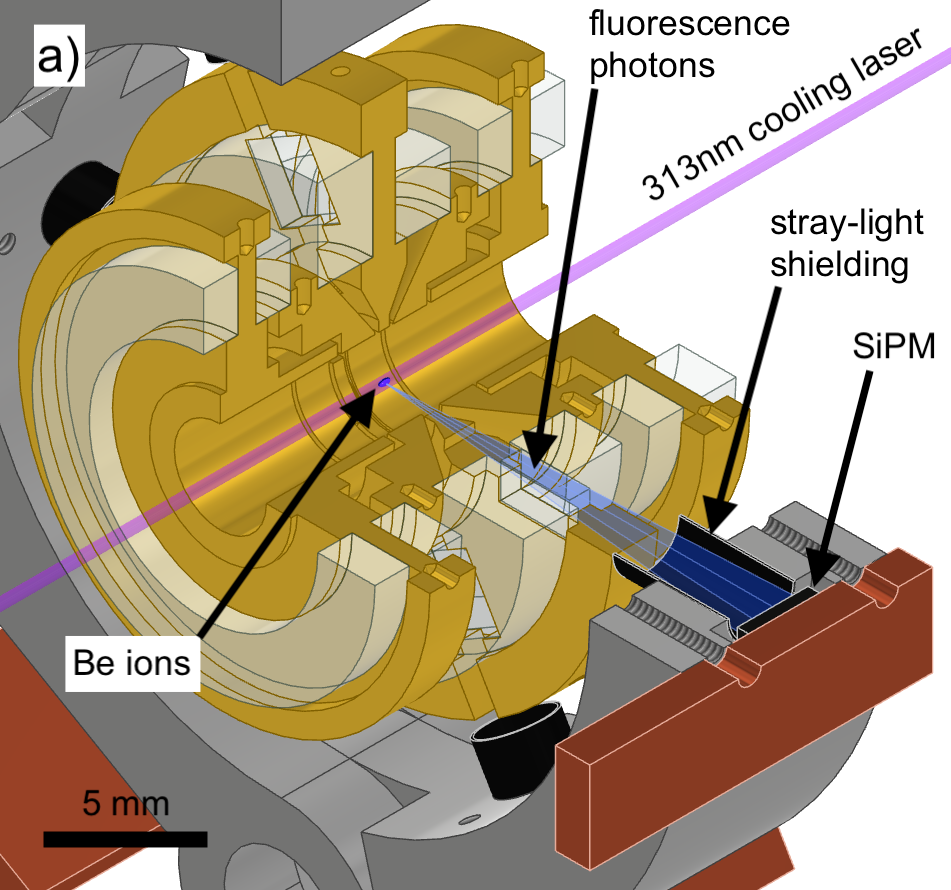}
    \includegraphics[width = 0.43\linewidth]{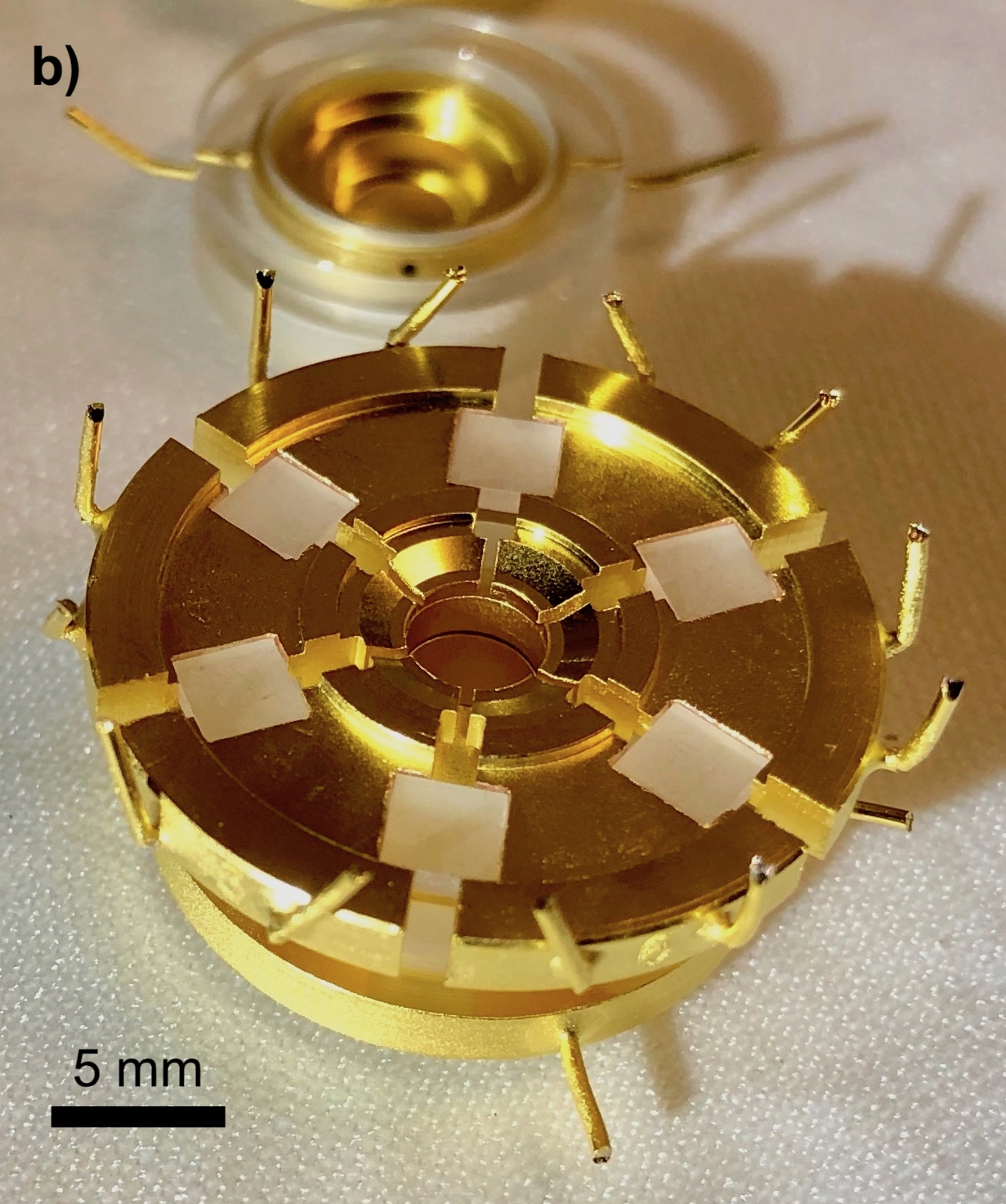}
    \caption{Section view of the BT setup (a). The solid angle of detected fluorescence photons is shown in light blue. Photograph of the BT electrodes (b). The two topmost electrodes have been removed to show the segmented ring electrode. Note that the sapphire blocks are polished only on the faces through which fluorescence light passes.}
    \label{fig:trap_setup} 
\end{figure}

Trap-integrated detection of fluorescence light is performed in the beryllium trap (BT) where laser-cooled \textsuperscript{9}Be\textsuperscript{+} ions are stored. This trap is a cylindrical open-endcap five-pole Penning trap with \SI{4}{\milli\meter} inner diameter designed to be orthogonal and compensated~\cite{Gabrielse1989}. A crucial additional feature of the BT is the six-fold segmented ring electrode shown in Fig.~\ref{fig:trap_setup}. The benefits are two-fold: first, it allows for the application of rotating-wall drives~\cite{Huang1997,Huang1998} to radially compress the stored ion cloud, and second, it allows scattered fluorescence photons from the \textsuperscript{9}Be\textsuperscript{+} ion cloud to escape the trapping volume. The slits between the electrode segments cover \ang{6} in azimuth angle, and \SI{0.785}{\milli\meter} in axial direction. Each slit allows about 0.3\,\% of the fluorescence light to escape the trap. About 0.087(17)\,\% of the overall fluorescence light can reach a single SiPM. The trap electrodes are made from gold-plated oxygen-free electrolytic (OFE) copper and are electrically isolated with sapphire rings. The six segments of the ring electrode are held in place by optically polished sapphire blocks.
A tube made of black anodized aluminum mounted in the holder next to the SiPM suppresses stray light from directions other than the center of the trap.
In addition, tubes with UV-absorbent coating are placed at the top and bottom of the trap stack for stray-light shielding, clipping the laser beam such that it does not hit the gold-plated electrodes.

Fluorescence photons from the trapping region pass the sapphire blocks and are detected by up to two SiPM mounted outside two of the six slits of the BT ring electrode. The distance from the SiPM detectors to the trapping region is approximately \SI{17}{\milli\meter}. Each SiPM is read out individually in photon counting mode.
The SiPM model MicroFJ-30035-TSV from \textit{onsemi} has been selected because it features a glass window, which makes the device more sensitive to the ultraviolet light of the \textsuperscript{9}Be\textsuperscript{+} laser-cooling transition near \SI{313}{\nano\meter}, and because a similar model from the same manufacturer was operated at \SI{4}{\kelvin} in previous work~\cite{Biroth2016}.
According to the data sheet, at \SI{313}{\nano\meter}, the photon detection efficiency (PDE) of the SiPM is 23\,\% at \SI{2.5}{\volt} overvoltage and 28\,\% at \SI{6.0}{\volt} overvoltage when operated at room temperature~\cite{SiPM_datasheet}.
The SiPM features an active area of $3 \times \SI{3}{\milli\meter\squared}$ covered by a total of 5676 microcells, each \SI{35}{\micro\meter} in size. The fill factor is 75\,\%.
Due to its insensitivity to magnetic fields, the SiPM is able to operate in the \SI{1.9}{\tesla} magnetic field of our Penning-trap system.
The power consumption of the SiPM depends on the count rate and is on the order of \SI{1}{\micro\watt} at room temperature and much lower at \SI{4}{\kelvin} due to the reduced dark count rate.

\begin{figure}
    \includegraphics[width = \linewidth]{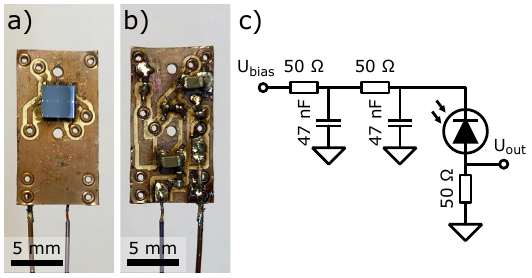}
    \caption{Photograph of the SiPM biasing and readout board front (a) and back (b). SiPM biasing and readout board circuit diagram (c).}
    \label{fig:SiPM_board}
\end{figure}

Each SiPM is soldered onto a small biasing and readout board which contains low-pass filters for the biasing voltage and a \SI{50}{\ohm} output resistance, as shown in Fig.~\ref{fig:SiPM_board}.
The board material \textit{Rogers} RO4350B has a low dielectric loss tangent and is suitable for cryogenic operation. 
The cabling from room temperature to \SI{4}{\kelvin} requires a compromise between low thermal conductivity to avoid excessive heat load to the cryogenic experiment and high signal transmission up to frequencies of approximately \SI{1}{\giga\hertz}. For the readout cable an \SI{0.51}{\milli\meter}-diameter semi-rigid coaxial cable of type PE-020SR from \textit{Pasternack} has been chosen. The small diameter suppressed heat flow while the silver plating of the inner conductor provides sufficient signal transmission. Using a \SI{1}{\meter}-long cable, which is thermally anchored at the liquid nitrogen stage of the cryostat, keeps the heat load to the 4-K stage below \SI{10}{\milli\watt}.
Two \textit{Mini-Circuits} ZFL-1000LN+ low-noise amplifiers mounted directly onto the SMA vacuum-feedthrough are used to amplify the signal before it is recorded with an oscilloscope, waveform digitizer or photon counter.
To supply the biasing voltage to the SiPM, \SI{0.05}{\milli\meter}-diameter manganin wires are used.

The cooling laser is a commercial TA-FHG pro diode laser system from \textit{Toptica}. An external cavity diode laser generates light near \SI{1252}{\nano\meter} which is amplified in a tapered amplifier and frequency doubled twice in two cascaded second harmonic generation (SHG) cavities.
The frequency is stabilized with a WSU8-2 wavelength meter from \textit{HighFinesse} using light near \SI{626}{\nano\meter} coupled out after the first SHG stage.
The \SI{313}{\nano\meter} light is transferred from the optical table to the magnet via a hydrogen-loaded single-mode photonic crystal fiber~\cite{Colombe2014}.
An optical breadboard bolted to the magnet below the entrance window to the horizontal bore hosts the beam delivery optics. The beam coming from the fiber is collimated and then polarized by an alpha-BBO Glan-laser polarizer. The rejected light from the polarizer is used to monitor the power of the \SI{313}{\nano\meter} laser light delivered to the experiment.
The polarization of the beam directed into the trap is adjusted using motorized half-wave and quarter-wave plates.
The beam position and angle are adjusted using a pair of motorized mirrors in front of the entrance window.

\section{SiPM characterization at room temperature and at 4\,K}\label{sec:SiPMcharacter}

\subsection{Cryocooler-based test setup}

\begin{figure}
    \includegraphics[width = \linewidth]{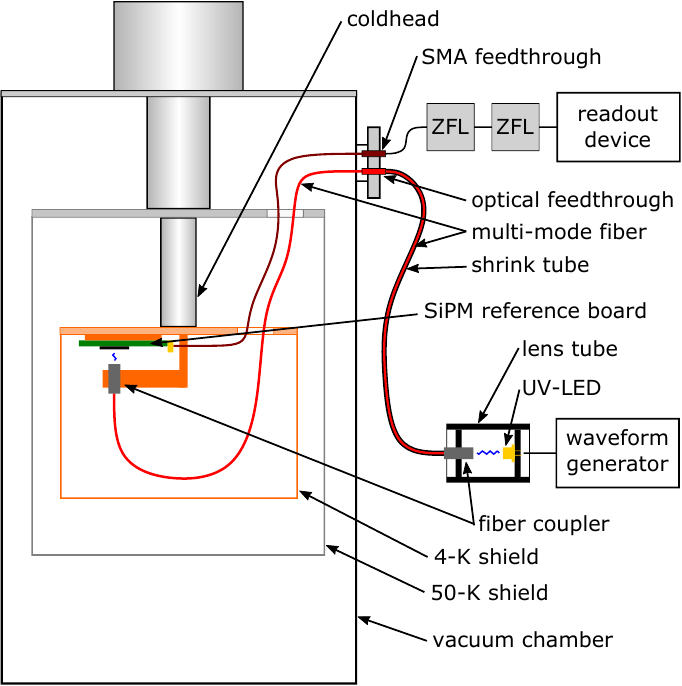}
    \caption{Cryocooler-based test setup for SiPM characterization at cryogenic (\SI{4}{\kelvin}) and room temperature. Cables for SiPM biasing are omitted. Readout device: either oscilloscope, waveform digitizer, or photon counter. ZFL: low-noise amplifier ZFL-1000LN+ from \textit{Mini-Circuits}.}
    \label{fig:test_setup}
\end{figure}

A cryocooler-based test setup is used to characterize and compare the properties of the SiPM at room temperature and at \SI{4}{\kelvin}.
For these measurements, a MicroFJ-SMA-30035 evaluation board, containing the MicroFJ-30035-TSV SiPM and its biasing and readout circuitry, is mounted to the 4-K stage of the pulse-tube cryo\-cooler. The 4-K section of the cryocooler is completely enclosed by a copper heat shield kept at \SI{4}{\kelvin} in order to eliminate heat load on the evaluation board due to thermal radiation. A second aluminum heat shield mounted to the 50-K stage of the cryo\-cooler reduces the heat load to the 4-K heat shield. A schematic of the setup is shown in Fig.~\ref{fig:test_setup}.
Two Cernox thin-film resistance temperature sensors are mounted on the 4-K stage for temperature measurements.
The cabling for biasing and readout of the SiPM evaluation board is the same as in the Penning-trap setup described above. The cables are thermally anchored at the 4-K and 50-K stages of the cryo\-cooler to avoid heat load on the evaluation board due to thermal conduction through the cables.

Light pulses are delivered to the SiPM through a multi-mode fiber. One end of the fiber is mounted to the 4-K stage at a distance of approximately \SI{10}{\milli\meter} from the SiPM. The fiber is routed outside the vacuum chamber using a fiber feedthrough.
A LED315W ultraviolet light-emitting diode (UV-LED) from \textit{Thorlabs} with emission around \SI{315}{\nano\meter} is used to generate short pulses of light containing only a few photons which are coupled into the other end of the fiber. 
The UV-LED is operated by applying rectangular pulses with a fixed pulse length of \SI{20}{\nano\second} and varying voltage and repetition rate from a waveform generator.
Care was taken to install the UV-LED and the fiber coupler inside a lens tube in a light-tight way. 
The section of the fiber outside of the vacuum chamber had to be enclosed in light-tight black shrink tubing in order to suppress light entering the fiber from light sources in the laboratory.
Light-tightness of the setup is checked by varying the brightness of these light sources and utilizing the extremely low dark count rate of the SiPM at \SI{4}{\kelvin} which allows to detect stray-light-photon count rates as low as \SI{1}{\per\second}.

Our test setup allows us to cool down the SiPM while keeping the single-photon source at a constant room temperature. As a consequence, temperature-dependent effects in the source are irrelevant, and the number of photons delivered to the SiPM is independent of temperature. This enables a direct comparison of the detection efficiency at room temperature and at \SI{4}{\kelvin}.

\subsection{Pulse shape}

\begin{figure}
    \includegraphics[width = 0.9\linewidth]{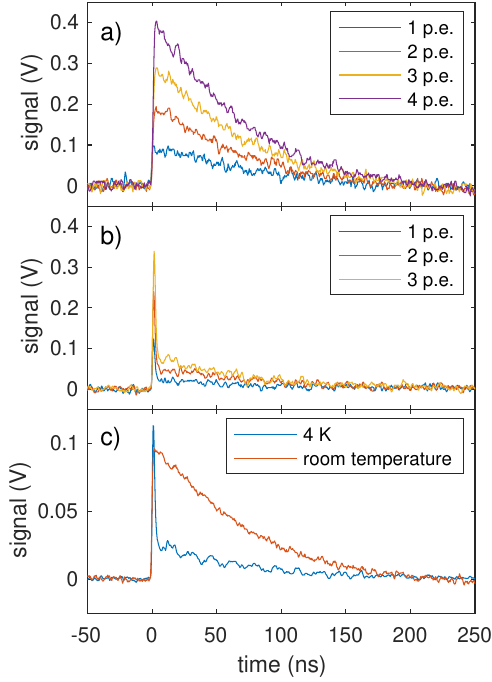}
    \caption{(a) Oscilloscope traces of SiPM pulses at room temperature and bias voltage \SI{26.5}{\volt}. (b) Oscilloscope traces at \SI{4}{\kelvin} and bias voltage \SI{24.0}{\volt}. In (a) and (b) one graph for each $n$-photoelectron pulse is shown. (c) Direct comparison of the SiPM pulse shape at \SI{4}{\kelvin} and room temperature. The average of 27 one-photoelectron pulses is plotted for both temperatures. p.e.: photoelectron.}
    \label{fig:pulses} 
\end{figure}

The output signal of the SiPM is the sum of the contributions from all microcells. The signal is therefore quantized with respect to the number of avalanching microcells and is a multiple of the signal of the one-photoelectron pulse. Graphs of such multi-photoelectron pulses are shown in Fig.~\ref{fig:pulses}\,a) for room temperature and in Fig.~\ref{fig:pulses}\,b) for \SI{4}{\kelvin}.
The shape of the one-photoelectron pulse at room temperature and at \SI{4}{\kelvin} is compared in Fig.~\ref{fig:pulses}\,c).

The typical pulse shape at room temperature is characterized by a fast rise with a rise-time on the order of \SI{1}{\nano\second} and an exponential decay. The time constant of the exponential decay is determined by the microcell recharge time constant $\tau_{RC} = R_q C_d$, where $R_q$ is the quench resistance and $C_d$ is the effective microcell capacitance~\cite{SiPM_whitepaper}. From a fit to the exponential decay we determine $\tau_{RC} = \SI{70.1(5)}{\nano\second}$.

At cryogenic temperature the fast rise is unchanged. However, the exponential decay is composed of two components. A fast component decaying with a time constant of \SI{1.9(4)}{\nano\second} to a level of about one quarter of the maximum and a slow component decaying with a time constant of \SI{74(1)}{\nano\second}.
Similar pulse shapes have been observed at cryogenic temperatures in Ref.~[\onlinecite{Otono2007}] and modelled in Ref.~[\onlinecite{Otono2009}]. The reason for the different pulse shape at cryogenic temperatures is an increased quench resistance. When the quench resistance becomes too large, the quenching occurs partially via the stray capacitance of the quench resistor instead, which explains the fast component.

To quantify the change in quench resistance of our SiPM, its value is calculated from the measured values of the recharge time constant $\tau_{RC}$ and the microcell capacitance $C_d$ evaluated in  section~\ref{sec:microcellc}.
The resulting values are listed in Tab.~\ref{tab:table1}. At cryogenic temperatures, we indeed observe that the quench resistance is increased. In addition, the microcell capacitance is reduced, while the recharge time constant shows only a minor change.
We attribute both the change in quench resistance and the change in microcell capacitance to temperature-dependent effects in silicon.

\begin{table}[b]
\caption{\label{tab:table1}
Measured values of recharge time constants $\tau_{RC}$, and microcell capacitances $C_d$, as well as calculated values of quench resistances $R_q$ for the MicroFJ-30035-TSV SiPM.}
\begin{ruledtabular}
\begin{tabular}{cccc}
 & $\tau_{RC}$ (ns) & $C_d$ (fF) & $R_q$ (M$\Omega$)\\
\hline
Room temperature  & 70.1(5) & 158(2) & 0.444(6) \\
$T \approx \SI{4}{\kelvin}$ & 74(1) & 35(2) & 2.1(1) \\
\end{tabular}
\end{ruledtabular}
\end{table}

\subsection{Charge and pulse height}

A SiPM pulse is characterized by two measures: its pulse height and its charge. The pulse height is defined as the maximum amplitude of the pulse with respect to the baseline. The charge $Q$ of the pulse is defined as the numerical integral over the pulse waveform.
\begin{equation} 
    Q = \frac{1}{G_A R} \int V(t) \mathop{\mathrm{d}t} ,
\end{equation} where $G_A$ is the voltage gain of the ZFL-amplifier chain, and $R = \SI{25}{\ohm}$ (the \SI{50}{\ohm} output resistance of the SiPM biasing and readout circuit in parallel to the \SI{50}{\ohm} impedance of the transmission line).

The baseline of the pulse is defined as the mean of the signal level in the time window ranging from \SI{1000}{\nano\second} to \SI{10}{\nano\second} before the trigger, and is determined for each pulse individually in order to take into account baseline fluctuations. Traces containing dark-count pulses in this time window are excluded from the analysis. For the subsequent determination of pulse height and charge, the baseline is subtracted from the signal level.
The data in the time window from \SI{10}{\nano\second} before the trigger to \SI{200}{\nano\second} after the trigger are then used to calculate the pulse height and the charge of an individual pulse.

\begin{figure}
    \includegraphics[width = \linewidth]{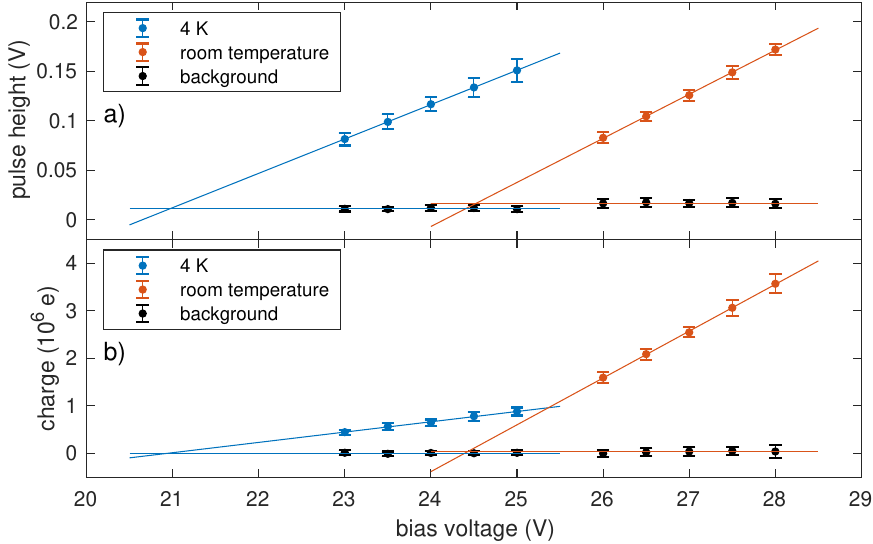}
    \caption{Pulse height (a) and charge (b) of one-photoelectron SiPM pulse waveforms as a function of bias voltage. Each datapoint is the average of approximately 50 waveforms, the errorbars indicate the 1-$\sigma$ standard deviation.}
    \label{fig:pulse_PH_charge_biasvoltage}
\end{figure}

In the following, we characterize the dependence of pulse height and charge on the bias voltage by analyzing oscilloscope traces of SiPM pulse waveforms.
For simplicity, we consider only one-photoelectron waveforms. 
The resulting values are shown in Fig.~\ref{fig:pulse_PH_charge_biasvoltage} where, in each panel, measurements at room temperature and at \SI{4}{\kelvin} are compared.
We observe a linear dependence of both pulse height and charge on bias voltage for both temperatures.

From a linear fit to the data in Fig.~\ref{fig:pulse_PH_charge_biasvoltage}\,a) we determine the dependence of pulse height on bias voltage to \SI{0.0347(2)}{\volt}/V at \SI{4}{\kelvin}. This is 22\,\% lower compared to the value at room temperature of \SI{0.0445(5)}{\volt}/V.
Based on a linear fit to the data in Fig.~\ref{fig:pulse_PH_charge_biasvoltage}\,b) we find that the dependence of charge on bias voltage is $0.217(11)\times10^6\,e$/V at \SI{4}{\kelvin}. This is a reduction by a factor of 4.5 compared to the value at room temperature of $0.986(13)\times10^6\,e$/V.

\subsection{Breakdown voltage, microcell capacitance, and gain}\label{sec:microcellc}

The breakdown voltage $U_0$ of the SiPM is determined by a linear extrapolation of the pulse height and charge to zero.
Since noise is superimposed onto the SiPM pulse, the measured pulse height and charge are modified, which needs to be taken into account.
We evaluate noise with the same algorithms as used for the evaluation of SiPM pulses, and obtain the background values shown in Fig.~\ref{fig:pulse_PH_charge_biasvoltage}. A finite value for the background pulse height is determined, while the value for the background charge is consistent with zero. Note that the noise pulse height differs between room temperature and \SI{4}{\kelvin}.
Consequently, we extrapolate the pulse height to the value given by the background pulse height and the charge to zero. 
The resulting extrapolations are shown in Fig.~\ref{fig:pulse_PH_charge_biasvoltage} as well. The estimates of the breakdown voltage based on pulse height and based on charge agree within the uncertainty of the measurement, and the resulting combined values are $U_0 = $ \SI{24.5(1)}{\volt} at room temperature and $U_0 = $ \SI{21.0(1)}{\volt} at \SI{4}{\kelvin}. Furthermore, the determined breakdown voltage at room temperature is in agreement with the value given in the data sheet~\cite{SiPM_datasheet}.
 
The microcell capacitance $C_d$ is determined by the slope of the charge $Q_1$ of a one-photoelectron pulse as a function of bias voltage since it is defined as
\begin{equation}
    C_d = \frac{Q_1}{\Delta U} = \frac{Q_1}{U - U_0} ,
\end{equation}
where $\Delta U = U - U_0$ is the overvoltage. 
A linear fit to the data in Fig.~\ref{fig:pulse_PH_charge_biasvoltage}\,b) gives a microcell capacitance of $C_d = \SI{158(2)}{\femto\farad}$ at room temperature and $C_d = \SI{35(2)}{\femto\farad}$ at \SI{4}{\kelvin}.
Compared to room temperature, the microcell capacitance is reduced by a factor of 4.5 at \SI{4}{\kelvin}.

The gain $G$ of the SiPM is determined by the relationship
\begin{equation}
    G = \frac{Q_1}{e} ,
\end{equation}
where $e$ is the elementary charge, and $Q_1$ is the charge of a one-photoelectron pulse.
The gain measured at room temperature is consistent with the values given in the data sheet~\cite{SiPM_datasheet}. Since the gain is proportional to the microcell capacitance it is also reduced by a factor of 4.5 at \SI{4}{\kelvin}.

\subsection{Crosstalk}

\begin{figure}
    \includegraphics[width = 0.75\linewidth]{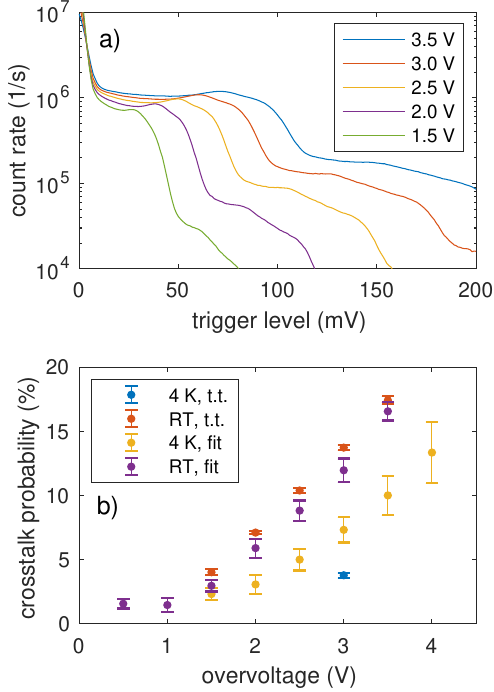}
    \caption{Dark count rate at room temperature (RT) as a function of trigger level for various overvoltages (a) and crosstalk probability as a function of overvoltage (b). t.t.: trigger-threshold method. fit: fit method.}
    \label{fig:crosstalk} 
\end{figure}

The crosstalk probability $q$ is the probability that a triggered microcell causes an additional and simultaneous avalanche in another microcell.
This probability can be determined based on a measurement of the dark count rate as a function of the trigger threshold.
For dark counts, the ratio of the count rates of two-photoelectron pulses to one-photoelectron pulses is an estimate of the crosstalk probability. 
 For this measurement, the SiPM is installed in the Penning-trap setup, and a SR400 photon counter from \textit{SRS} is used to record the count rate. Stray-light is suppressed, such that dark counts dominate. 
 The recorded dark count rate at room temperature is shown in Fig.~\ref{fig:crosstalk}\,a) for various overvoltages. The resulting crosstalk probability is shown in Fig.~\ref{fig:crosstalk}\,b). The data show the typical increase of the crosstalk probability with overvoltage.

\begin{figure}
    \includegraphics[width = 0.75\linewidth]{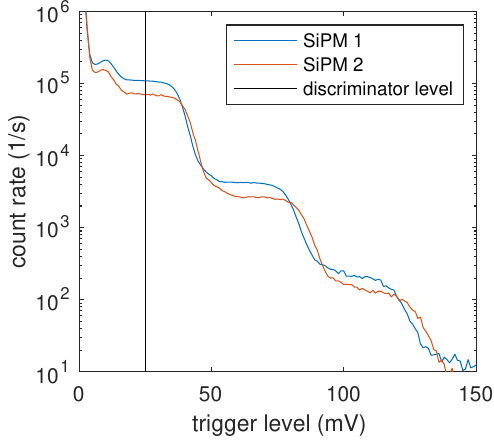}
    \caption{Count rate at \SI{4}{\kelvin} as a function of trigger level for two SiPM installed in the BT and biased with \SI{24.0}{\volt}. The data is recorded with a SR400 photon counter. The threshold chosen to discriminate background and one-photoelectron pulses is marked with a black vertical line at \SI{25}{\milli\volt}. Note that, in order to speed up this measurement, additional light was introduced into the trap to compensate for the low dark count rate at \SI{4}{\kelvin}.}
    \label{fig:counts_over_threshold}
\end{figure}

At \SI{4}{\kelvin}, the dark count rate is too low to determine the crosstalk probability based on dark counts. Instead, fluorescence light from \textsuperscript{9}Be\textsuperscript{+} ions is used. The fluorescence light level is chosen so low that the probability of two photons arriving at the same time is negligible. The recorded count rate is shown in Fig.~\ref{fig:counts_over_threshold} and the resulting crosstalk probability in Fig.~\ref{fig:crosstalk}\,b). For the typical bias voltage of $U = \SI{24.0}{\volt}$ used at \SI{4}{\kelvin}, the crosstalk probability is 3.8(2)\,\%. This is a factor of three lower than at room temperature at the same overvoltage.

In addition to the trigger-threshold method described above, the crosstalk probability is  also determined from a fit to the photoelectron distribution, as introduced in the next section. For this measurement, the SiPM is installed in the cryo\-cooler-based test setup and is read out by a waveform digitizer. The values resulting from the fit are shown in Fig.~\ref{fig:crosstalk}\,b) as well.
This method gives a crosstalk probability at \SI{4}{\kelvin} which is about a factor of two lower than at room temperature. 

Overall, the crosstalk probability at \SI{4}{\kelvin} is significantly reduced compared to room temperature.
At room temperature, the values from both methods show only small deviations.
However, at \SI{4}{\kelvin} the trigger-threshold method results in a factor of 2 lower estimate than the fit method.
The discrepancies might be explained by the different processes which are used to trigger the microcells. For the trigger-threshold method dark counts are used at room temperature and \SI{313}{\nano\meter} fluorescence photons at \SI{4}{\kelvin}, while for the fit method UV-LED light pulses near \SI{315}{\nano\meter} are used at both room temperature and \SI{4}{\kelvin}. The trigger-threshold measurements and fit measurements have been performed using different SiPM in different environments, so that the discrepancy may also arise from batch variation or the environmental conditions.

\subsection{Photon detection efficiency}\label{sec:pde}

To characterize the photon detection efficiency (PDE) near \SI{313}{\nano\meter}, UV-LED light pulses containing only a few photons are applied to the SiPM installed in the cryocooler-based test setup. 
Subsequently, the mean number of detected photons $\lambda$ per UV-LED light pulse is determined from photoelectron distributions. Finally, a relation between $\lambda$ and the PDE is established by comparing $\lambda$ with known values of the PDE~\cite{SiPM_datasheet}.

Here, we record SiPM pulse waveforms using a DT 5761 waveform digitizer from \textit{CAEN} which is triggered synchronously with the applied UV-LED light pulses. A repetition rate of \SI{10}{\kilo\hertz} assures suppression of accidental recordings of afterpulses and dark counts.
All synchronous responses of the SiPM to UV-LED light pulses are recorded, including waveforms that generate a zero-photoelectron response on the SiPM.

\begin{figure}
    \includegraphics[width = 0.75\linewidth]{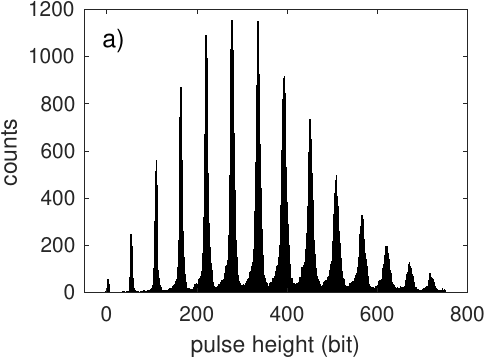}
    \includegraphics[width = 0.75\linewidth]{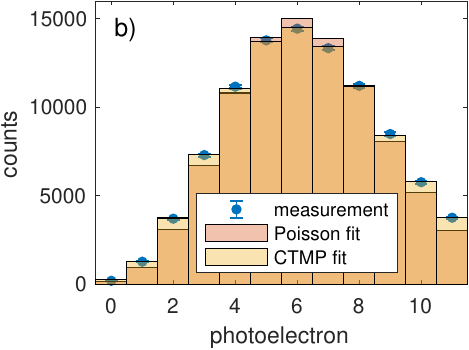}
    \caption{(a) SiPM pulse-height distribution for a bias voltage of \SI{27.0}{\volt} at room temperature and UV-LED light intensity setting 1. (b) The photoelectron distribution resulting from \ref{fig:pulse_height_dist_at_RT}a is fitted with a Poisson distribution and a crosstalk-modified Poisson (CTMP) distribution. The better fit is achieved by the CTMP distribution with $\lambda = 5.96(2)$ and $q = 0.081(4)$.}
    \label{fig:pulse_height_dist_at_RT}
\end{figure}

The baseline-compensated pulse-height distribution from such a measurement is shown in Fig.~\ref{fig:pulse_height_dist_at_RT}\,a) for room temperature, a SiPM bias voltage of \SI{27.0}{\volt}, and UV-LED light intensity setting 1. The peaks in the pulse-height distribution correspond to $n$-photoelectron pulses. In order to improve the resolution of these peaks, a \SI{22}{\mega\hertz} low-pass filter (SLP-21.4+ from \textit{Mini-Circuits}) has been installed at the input of the waveform digitizer. This slightly distorts the pulse shape but increases the resolving power of the individual peaks considerably.
For further evaluation, all counts within the corresponding peaks of the pulse-height distribution are summed up, resulting in the photoelectron distribution shown in Fig.~\ref{fig:pulse_height_dist_at_RT}\,b).

The UV-LED light source can be described as a thermal light source with a Poissonian photon distribution. However, crosstalk modifies the measured photoelectron distribution, since for each avalanching microcell an additional microcell is triggered with crosstalk probability $q$.
This effect is taken into account using a crosstalk-modified Poisson (CTMP) distribution~\cite{Biroth2015} with parameters $\lambda$ and $q$. For $q \to 0$ this distribution converges to the Poisson distribution with parameter $\lambda$. 
We fit one of the photoelectron distributions with both a Poisson distribution and the CTMP distribution, and compare the results in Fig.~\ref{fig:pulse_height_dist_at_RT}\,b). While the Poisson distribution systematically deviates from the measured data, the data is well described by the CTMP distribution. The CTMP distribution further allows to extract independent values for the crosstalk probability $q$, shown in Fig.~\ref{fig:crosstalk}\,b) as a function of overvoltage.

\begin{figure}
    \includegraphics[width = \linewidth]{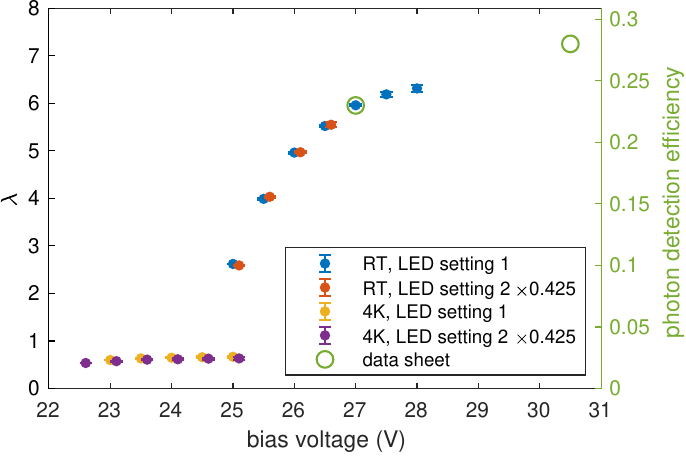}
    \caption{Mean number of detected photons $\lambda$ and photon detection efficiency (PDE) as a function of bias voltage, at room temperature (RT) and \SI{4}{\kelvin}, and for two settings of the UV-LED light intensity. The measurements for setting 2 have been scaled by a factor of 0.425 and shifted by \SI{0.1}{\volt} for better visualization. The value at RT and \SI{27.0}{\volt} bias voltage is used to calibrate the PDE to $\lambda$.}
    \label{fig:relative_PDE}
\end{figure}

The mean number of detected photons $\lambda$ from fits to photoelectron distributions is plotted in Fig.~\ref{fig:relative_PDE} for two UV-LED light intensity settings, with the SiPM at room temperature and \SI{4}{\kelvin}, and as a function of bias voltage.
The graph shows that $\lambda$ increases with bias voltage at room temperature.
At \SI{4}{\kelvin}, the dependence on bias voltage is reduced and $\lambda$ is smaller by a factor of 5 to 10.

The PDE of the SiPM is shown on the vertical axis on the right in Fig.~\ref{fig:relative_PDE}. 
It has been calibrated by relating $\lambda$ to the PDE at \SI{313}{\nano\meter} of 23\,\%, given in the data sheet~\cite{SiPM_datasheet} for room temperature and an overvoltage of \SI{2.5}{\volt}. 
Since the number of applied photons only depends on the UV-LED setting, this calibration is valid for all bias voltages and both temperatures, and establishes a relation between $\lambda$ and the PDE.
Two calibrations based on two different UV-LED light intensity settings agree.
For the bias voltage of \SI{24.0}{\volt}, typically used in the Penning-trap setup at \SI{4}{\kelvin}, we determine a PDE of 2.5(3)\,\%.

\section{Trap-integrated detection of $^9\textnormal{Be}^+$ fluorescence}\label{sec:fluorescence_detection}

We demonstrate our SiPM-based detection method with a cloud of \textsuperscript{9}Be\textsuperscript{+} ions stored in the BT, whose axial oscillation frequency is brought into resonance with the LC circuit at \SI{4}{\kelvin}. The ion number $N$ is determined from the line shape of the frequency spectrum of the LC circuit \cite{Sellner2017}, and is $N = 540(40)$ in run 1 and $N=5100(200)$ in run 2. Circularly polarized laser light near \SI{313}{\nano\meter} with laser power between \SI{60}{\micro \watt} and \SI{1800}{\micro\watt} is used to cool the \textsuperscript{9}Be\textsuperscript{+} ions.
In the \SI{1.9}{\tesla} magnetic field of the BT, \textsuperscript{9}Be\textsuperscript{+} ions can be cooled either on the $^{2}\textnormal{S}_{1/2}$ ($m_\textnormal{J}$ = 1/2) $\rightarrow$ $^2\textnormal{P}_{3/2}$ ($m_\textnormal{J}$ = 3/2) transition using $\sigma^+$ polarized light or on the $^2\textnormal{S}_{1/2}$ ($m_\textnormal{J}$ = \num{-1}/2) $\rightarrow$ $^2\textnormal{P}_{3/2}$ ($m_\textnormal{J}$ = \num{-3}/2) transition using $\sigma^-$ polarized light. Both options are closed cycling transitions with an intrinsic off-resonant repumping mechanism \cite{Wineland1980,Itano1981}. Using pure circularly polarized light ensures a bright state population $> \SI{99}{\percent}$.
We have cooled \textsuperscript{9}Be\textsuperscript{+} ions on and observed fluorescence signals for both transitions using appropriately polarized laser light.
In the following, we use the $^2\textnormal{S}_{1/2}$ ($m_\textnormal{J}$ = \num{-1}/2) $\rightarrow$ $^2\textnormal{P}_{3/2}$ ($m_\textnormal{J}$ = \num{-3}/2) transition. 

The SiPM is operated with a bias voltage of \SI{24.0}{\volt} and SiPM pulses are counted on the SR400 photon counter set to a trigger threshold of \SI{25}{\milli\volt} and a counting window of \SI{1000}{\milli\second}.
The ideal trigger threshold to discriminate one-photoelectron pulses from the noise was determined from a measurement of the background count rate as a function of threshold, as shown in Fig.~\ref{fig:counts_over_threshold}.

\begin{figure}
    \includegraphics[width = \linewidth]{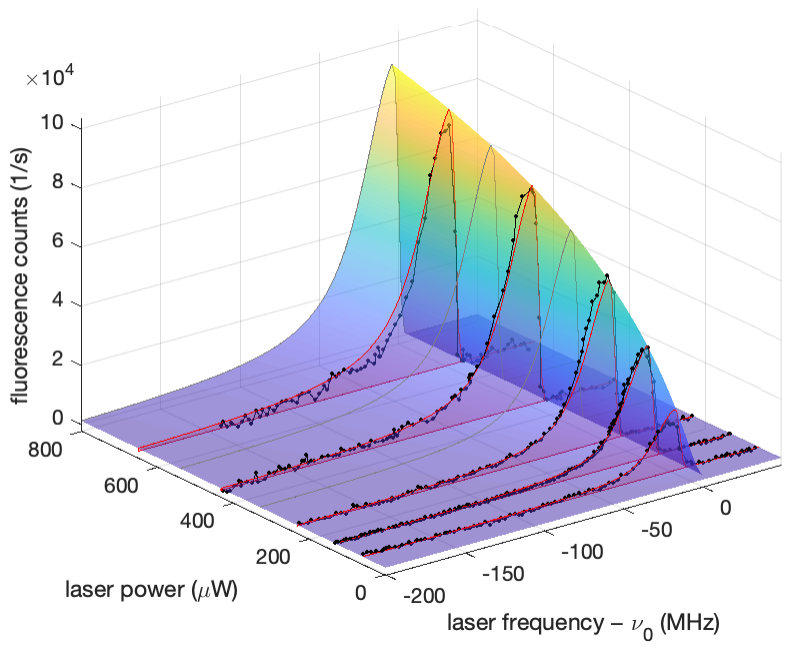}
    \caption{Fluorescence spectra of run 1. Black data points show the count rate of fluorescence photons as a function of the cooling-laser frequency and power. The laser frequency is scanned across the resonance from low to high frequencies for different values of the laser power. The sharp drop in fluorescence counts is caused by heating the ions out of resonance when the laser detuning becomes positive. Red curves show the result of the 2-dimensional fit at the laser power of the frequency scans. The color-coded surface shows the 2-dimensional fit, color-coded with respect to the fluorescence count rate. Grey line profiles are added to guide the eye.}
    \label{fig:fluorescenceinexp}
\end{figure}

We scan the laser frequency across the resonance from low to high frequencies with a scan rate of \SI{2}{\mega\hertz\per\second} and record the count rate of fluorescence photons. These scans are repeated for several values of laser power. The resulting background-removed data are shown in Fig.~\ref{fig:fluorescenceinexp}.
The fluorescence signal slowly rises with increasing laser frequency and follows a Voigt line profile. At the moment the laser frequency reaches the resonance frequency of the cooling transition, the fluorescence intensity sharply drops to zero as the ions are heated out of resonance. We further observe power broadening of the linewidth and saturation of the fluorescence count rate with increasing laser power.

The line shape of the fluorescence count rate is modelled as a Voigt profile $V(\nu,P)$ which is cut off at the resonance frequency $\nu_0$. The Voigt profile is the convolution of a Lorentzian profile $L(\nu,P)$ and a Gaussian distribution $G(\nu)$ with standard deviation $\sigma$.
The Lorentzian profile is defined as
\begin{equation}\label{eq:line_profile}
    L(\nu,P) = \eta I_C \frac{(\gamma/2)^2}{(\gamma/2)^2+(\nu-\nu_0)^2}
\end{equation}
with the power-broadened line width (FWHM) $\gamma = \gamma_0 \sqrt{1+P/P_0}$, the on-resonance scattering rate $I_C = \frac{2 \pi \gamma_0}{2} \frac{P/P_0}{1+P/P_0}$, the natural linewidth (FWHM) $\gamma_0 = \SI{19.6(10)}{\mega\hertz}$ \cite{Andersen1969}, the saturation power $P_0$, and the laser power $P$.
The parameter $\eta$ is the product of the total detection efficiency and the ion number, expressing the count rate of detected photons in terms of the saturated on-resonance scattering rate of a single ion. 
While power broadening and saturation is included in the Lorentzian part of the Voigt profile, Doppler broadening and other broadening effects are included in the Gaussian width of the Voigt profile.

\begin{figure}
    \includegraphics[width = \linewidth]{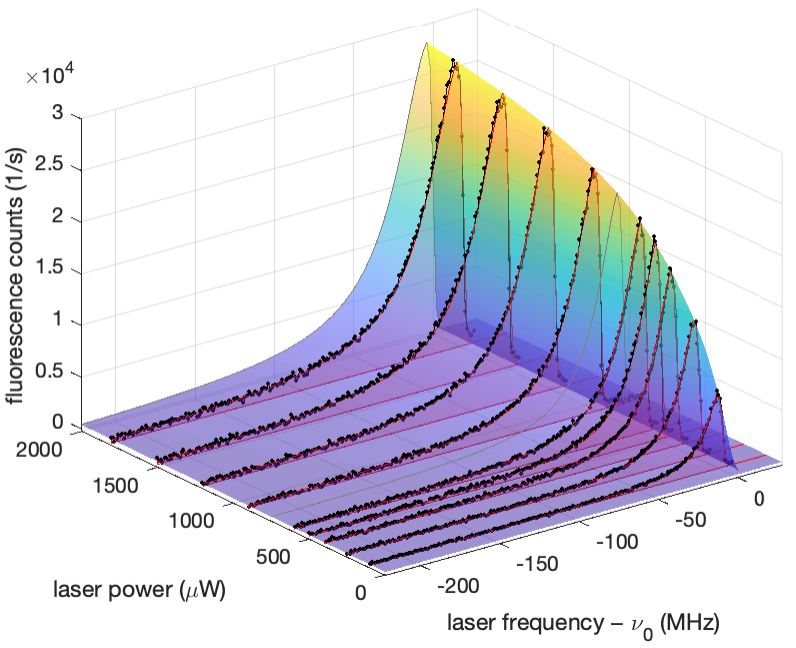}
    \caption{Fluorescence spectra of run 2. Description as in Fig.~\ref{fig:fluorescenceinexp}. In run 2, we observe a reduced background count rate of \SIrange[range-units=single]{6}{7}{\per\second} per \si{\micro\watt}, compared to \SIrange[range-units=single]{250}{1300}{\per\second} per \si{\micro\watt} in run 1, due to stray light suppression. In addition, stabilization of the laser power in run 2 leads to a background count rate which is independent of the laser frequency. Further, the laser frequency is stabilized to \SI{2}{\mega\hertz} peak-to-peak fluctuation, compared to approximately \SI{20}{\mega\hertz} in run 1. Finally, in run~2, no RF drive is applied, which eliminates line-broadening effects and allows for a better temperature estimate.}
    \label{fig:fluorescenceinexp_run2}
\end{figure}

First, the resonance curves for each laser power are fitted individually with the Voigt line profile added to a linear background (in run 1) or to a constant background (in run 2) to determine the background count rate.
The resulting background-removed data are shown in Fig.~\ref{fig:fluorescenceinexp} and~\ref{fig:fluorescenceinexp_run2}. Note that the laser power in run 2 is stabilized to better than 0.3\,\% while in run 1 the laser power fluctuates and drifts up to 10\,\% during a scan. The background-removed data are then simultaneously fitted with the Voigt line profile as a function of frequency and laser power. This 2-dimensional fit simultaneously accounts for power broadening and saturation which both depend on the ratio $P/P_0$. 
The resulting fit surface is plotted in Fig.~\ref{fig:fluorescenceinexp} and~\ref{fig:fluorescenceinexp_run2} as well. The fit parameters for run 1 are $\eta = 0.00224(6)$, $P_0 = \SI{212(10)}{\micro\watt}$, and $\sigma = \SI{9.3(4)}{\mega\hertz}$.
For the approximately 10 times larger ion cloud in run 2 the fit parameters are $\eta = 0.000546(2)$, and $P_0 = \SI{326(2)}{\micro\watt}$. The fit parameter $\sigma$ of the 2-dimensional fit converges to zero, therefore, $\sigma = \SI{3.3(3)}{\mega\hertz}$ is determined from the weighted mean of the individual fits.

Considering equation~\ref{eq:line_profile} for $\nu = \nu_0$ and $P/P_0 \to \infty$, as would be the case for a saturated transition, the count rate of detected photons $n_d$ is maximum and becomes
\begin{equation}
    n_d = \eta_e  \eta_g  \eta_a \eta_d  N \frac{2 \pi \gamma_0}{2} = \eta \frac{2 \pi \gamma_0}{2}
\end{equation}
where $\eta_e = 0.75$ is a correction factor due to non-isotropic emission from $\sigma^\pm$ transitions in a magnetic field \cite{Itano1982}, $\eta_g = 0.00087(17)$ is the geometrical acceptance of the SiPM, $\eta_a = 0.84(2)$ takes into account the absorption in the sapphire blocks, $\eta_d$ is the detection efficiency of the SiPM, and $N$ is the ion number.
Under these conditions the ion cloud has a well-defined photon scattering rate $N \pi \gamma_0$. This photon source is then used to independently characterize the detection efficiency of the SiPM.
Taking the value of $\eta$ from the fit of run 1, we evaluate the detection efficiency of the SiPM to $\eta_d = \frac{\eta}{N \eta_e \eta_g \eta_a} = 0.0075(16)$. This value is a factor of $3.3(8)$ lower than the detection efficiency resulting from the characterization in the cryocooler of $\eta_d = 0.025(3)$.
The reduced detection efficiency might be explained by the effects of the \SI{1.9}{\tesla} magnetic field in the Penning trap, which is not present for the characterization measurements in the cryocooler.
In run 2 we evaluate the detection efficiency of the SiPM to $\eta_d = 0.00019(4)$, which is a factor of $128(31)$ smaller than the detection efficiency in the cryocooler and a factor of $40(12)$ smaller than in run 1.
After run 2 we observed cracks in the glass windows of some of the installed SiPM due to repeated cooling cycles.
Attenuation due to these cracks could explain the additional reduction in detection efficiency and the variation in detection efficiency between the two examples of SiPM. A misalignment of the SiPM with respect to the slits in the BT ring electrode, which would change the geometrical acceptance, is another possibility.
In run 3, with newly installed SiPM, we observed a detection efficiency comparable to run 1.
The best total detection efficiency of our SiPM-based detection method was achieved in run 1 where $\eta/N = 4.2(3)\times10^{-6}$.

The count rate of detected photons per ion is $n_1 = n_d /N  = \frac{P/P_0}{1+P/P_0} \times \SI{256(24)}{\per\second}$ on resonance in run 1. 
This count rate is to be discriminated from the background count rate $n_b = P/P_0 \times \SI{5e4}{\per\second}$ which is dominated by stray light and increases linearly with laser power. The dark count rate is independent of laser power and contributes less than \SI{1}{\per\second} to the background count rate.
Therefore, the signal-to-background ratio is maximum at low laser power and decreases as the transition is saturated at high laser power.
Assuming signal-to-background ratios $\le 1$ and considering counting statistics, the ion sensitivity, defined as the fluorescence count rate divided by the uncertainty of the total count rate, is maximum near $P/P_0 = 1$.
At this laser power the signal-to-background ratio for a small ion cloud with $N = 10$ is approximately 0.025, and the ion cloud can be discriminated from the background with five standard deviations within an averaging time of \SI{0.8}{\second}. For smaller ion clouds this time increases proportional to $1/N^2$. 
If the background count rate due to stray light can be eliminated, the background would be dominated by the dark count rate of the SiPM, and single-ion sensitivity can be achieved with averaging times below \SI{100}{\milli\second}.
Besides reducing stray light, the single-ion sensitivity can be improved by increasing the geometrical acceptance $\eta_g$, or by using a sensor with higher detection efficiency.

The temperature of the laser-cooled \textsuperscript{9}Be\textsuperscript{+} ions is determined from the Gaussian broadening of the Voigt line profile. The fit results in a Gaussian broadening of $\sigma = \SI{9.3(4)}{\mega\hertz}$ in run 1 and $\sigma = \SI{3.3(3)}{\mega\hertz}$ in run 2, which, for \textsuperscript{9}Be\textsuperscript{+} ions, corresponds to a temperature of \SI{9(1)}{\milli\kelvin} and \SI{1.1(2)}{\milli\kelvin}, respectively. The evaluated temperature in run 1 is significantly larger than the Doppler limit of \SI{0.5}{\milli\kelvin}, while in run 2 the evaluated temperature is close to the Doppler limit. In both cases, the ions are heated due to the coupling to the LC circuit which acts as a thermal bath at a temperature of \SI{4}{\kelvin}. In run 1, an additional radio-frequency (RF) drive was used for mode coupling, leading to broadening similar to micromotion-induced broadening in RF traps~\cite{Bluemel1989}. The temperature estimate above is derived assuming that thermal Doppler broadening is the only broadening effect. Therefore, in case there are other broadening effects present, the estimated temperature constitutes an upper limit for the ion temperature. Consequently, this result demonstrates our ability to cool \textsuperscript{9}Be\textsuperscript{+} ions to the low temperatures necessary for sympathetic cooling of protons for ultra-high precision $g$-factor measurements~\cite{Will2022}. 

As an additional consistency check, the beam radius at the position of the ions was measured to $w = \SI{268(2)}{\micro\meter}$. 
This allows us to relate the total power $P$ in our Gaussian beam to the intensity at the center $I$ as
\begin{equation}
    P = \frac{\pi}{2} w^2 I.
\end{equation}
Setting $I$ to the saturation intensity for \textsuperscript{9}Be\textsuperscript{+} of $I_0 = \SI{840(40)}{\watt\per\meter\squared}$ and taking into account anisotropic absorbtion for $\sigma^\pm$-transitions, we calculate the saturation power to $P_0 = \SI{63(3)}{\micro\watt}$. In the experiment, we observe saturation at $P = \SI{212(10)}{\micro\watt}$ in run 1 and $P = \SI{326(2)}{\micro\watt}$ in run 2, which is a factor of $3.4(2)$ and $5.2(3)$ higher than the estimate. This deviation is consistent with the ions being positioned off-center where the intensity is lower and higher power is necessary to achieve saturation. Positioning the ions off-center is necessary to create an intensity gradient across the ion cloud  which is necessary for cooling the radial modes~\cite{Itano1982}.

\section{Conclusions}\label{sec:conclusion}

We presented a SiPM-based fluorescence-detection system for use in our next-generation proton $g$-factor measurement setup, provided a detailed characterization of the SiPM properties at room temperature and at \SI{4}{\kelvin}, and demonstrated its applicability for the detection of fluorescence photons from laser-cooled \textsuperscript{9}Be\textsuperscript{+} ions stored in our cryogenic Penning-trap system. 

Fluorescence detection provides direct information about the cooling rate during Doppler cooling and the final temperature of the laser-cooled ions. 
This information is not accessible with the regularly used image current detection systems, especially for large cooling rates where the \textsuperscript{9}Be\textsuperscript{+} ions decouple from the LC circuit. 

The presented SiPM setup constitutes a compact cryogenic fluorescence-detection system, that eliminates the need for optical detection pathways into the hermetically-sealed cryogenic Penning-trap chamber.
This is a considerable advantage as this reduces the radiative heat load on the liquid helium stage and allows for a compact trap design.
A further appreciable advantage of our approach is the use of a low-cost and readily-available commercial SiPM sensor, avoiding the production of custom micro-fabricated devices.
For this reason, SiPM-based fluorescence detectors might be an attractive alternative to custom micro-fabricated superconducting sensors~\cite{Slichter2021} or custom chip-integrated avalanche photodiodes~\cite{Reens2022} in quantum information processing experiments in radio-frequency traps, for both cryogenic and room temperature experiments.

Characterizing the SiPM, we found that it can be reliably operated at \SI{4}{\kelvin}, and observed detection efficiencies of 2.5(3)\,\% in the cryocooler-based test setup and 0.75(16)\,\% in the experiment. We found dark count rates below \SI{1}{\per\second} for both cases.
The pulse shape is modified due to a reduced microcell capacitance and increased quench resistance at \SI{4}{\kelvin} which manifests in a reduced charge of the SiPM pulses while the pulse height is unchanged. Further, the breakdown voltage is reduced by \SI{3.5}{\volt} and the crosstalk probability is a factor of two to three smaller than at room temperature.

In the experiment, axial temperatures of the laser-cooled \textsuperscript{9}Be\textsuperscript{+} ion cloud as low as \SI{1.1(2)}{\milli\kelvin} have been observed with our trap-integrated fluorescence-detection system. Using such a laser-cooled \textsuperscript{9}Be\textsuperscript{+} ion cloud as cooling medium for the proton axial mode, e.g.~by energy exchange via a common-endcap electrode or shared LC circuit \cite{Will2022,Tu2021}, can potentially reduce the proton axial temperature by a factor of up to 4000, compared to state-of-the-art experiments~\cite{Schneider2017,Smorra2017, Borchert2022}. 

Regarding ion sensitivity, our fluorescence-detection system provides a total detection efficiency of $4.2(3)\times10^{-6}$, corresponding to a photon count rate of $\frac{P/P_0}{1+P/P_0} \times \SI{256(24)}{\per\second}$ per ion. This results in a fast detection of ion clouds with more than 10 ions with averaging times lower than \SI{1}{\second}. The ion sensitivity is predominantly limited by stray light. Therefore, additional stray light suppression measures, e.g.~focusing the fluorescence light through a narrow aperture onto the SiPM, can significantly improve the signal-to-background ratio and ion sensitivity. 
If a reduction by a factor 100 can be achieved, the system can be used to resolve fluorescence from a single \textsuperscript{9}Be\textsuperscript{+} ion within an averaging time shorter than \SI{1}{\second}.

Cooling of charged particles below the liquid helium temperature is becoming essential in various precision physics applications, e.g.~for precision measurements on the helion~\cite{Schneider2022}, highly-charged ions \cite{Tu2021}, and protons and antiprotons \cite{Schneider2017,Smorra2017,Borchert2022}. 
In particular, high-precision measurements of proton and antiproton $g$-factors require ultra-low temperatures for high-fidelity readout of the spin state \cite{Mooser2013}, and would immensely profit from the low temperatures reached with laser-cooled \textsuperscript{9}Be\textsuperscript{+} ions. Fluorescence-based detection using compact cryogenic SiPM detectors with the presented performance will facilitate sympathetic cooling by laser-cooled ions~\cite{Bohman2021,Will2022}, which will allow to cool single ions to temperatures in the \si{\milli\kelvin} regime in future multi-Penning trap experiments.

Further interesting applications for such a SiPM based detection system are fast non-destructive measurements of the motional frequencies of the trapped ion based on the detection of a reduced photon scattering rate due to the Doppler shift induced by a resonant excitation of the trapped ion motion~\cite{Wineland1983,Rodriguez2012}. 
Also, a two-ion crystal, in our case composed of a proton and a \textsuperscript{9}Be\textsuperscript{+} ion, would compose an interesting system for measurements of the motional frequencies or charge-to-mass ratios~\cite{Gutierrez2019b,Berrocal2022}.

Ultimately, using advanced laser-cooling techniques to bring a \textsuperscript{9}Be\textsuperscript{+} ion into the motional ground state, e.g.~by Raman sideband cooling~\cite{Heinzen1990} or EIT cooling~\cite{Morigi2000}, the presented SiPM-based detection system can be used to perform state readout for quantum-logic detection of the Larmor frequency and motional frequencies, either for co-trapped ions or coupled ions stored in separate traps~\cite{Heinzen1990,Will2022,Cornejo2021}.

\begin{acknowledgments}
This study comprises parts of the PhD thesis work of M.W. 
We thank the group of Prof. Achenbach for helpful advice on selecting a suitable SiPM model.
We acknowledge financial support from 
the Max-Planck Society, 
the RIKEN Chief Scientist Program, 
the European Union under the Horizon 2020 research and innovation programme (Marie Skłodowska-Curie grant agreement No 721559 - AVA, ERC grant agreement No 832848 - FunI, and ERC grant agreement No 852818 - STEP), 
and the Max-Planck–RIKEN–PTB Center for Time, Constants and Fundamental Symmetries.
\end{acknowledgments}

\section*{Data Availability Statement}
The data supporting the findings presented in this manuscript are available from the corresponding author upon reasonable request.

\bibliography{SiPM-refs}

\end{document}